\documentclass[runningheads,a4paper]{llncs}

\usepackage{amssymb}
\usepackage{graphicx}

\usepackage{amsmath}
\usepackage{hyperref}

\usepackage{algorithm}
\usepackage{algorithmic}

\usepackage{authblk}

\usepackage{subcaption}

\usepackage{cite}

\usepackage{url}
\urldef{\mailsa}\path|{bista@dmi.unipg.it, paolo.giuliodori@unicam.it}|
\urldef{\mailsb}\path|anna.kramer, leonie.kunz, christine.reiss, nicole.sator,|
\urldef{\mailsc}\path|erika.siebert-cole, peter.strasser, lncs}@springer.com|

\newtheorem{defn}{Definition}
\newtheorem{prop}{Proposition}

\begin{document}

\mainmatter  

\title{Mechanism Design Approach for Energy Efficiency}

\titlerunning{Mechanism Design Approach for Energy Efficiency}

%
%
\author{Stefano Bistarelli\inst{1}, Rosario Culmone\inst{2}, Paolo Giuliodori\inst{2} \and Stefano Mugnoz\inst{3}%
}
\authorrunning{Mechanism Design Approach for Energy Efficiency}

\institute{University of Perugia, Computer Science Department, Italy \and University of Camerino, Computer Science Department, Italy  \and U-Space SRL, Rome - Camerino, Italy
\mailsa\\
}


%
%

\toctitle{Mechanism Design Approach for Energy Efficiency}
\tocauthor{Authors' Instructions}
\maketitle

\begin{abstract}
In this work we deploy a mechanism design approach for allocating a divisible commodity (electricity in our example) among consumers. We consider each consumer with an associated personal valuation function of the energy resource during a certain time interval. We aim to select the optimal consumption profile for every user 
avoiding consumption peaks when the total required energy could exceed the energy production. The mechanism will be able to drive users in shifting energy consumptions in different hours of the day.
We start by presenting a very basic Vickrey-Clarke-Groves mechanism, we discuss its weakness and propose several more complex variants. 
\end{abstract}

\section{Introduction}
In this work, we investigate the trade-off between a final user's point of view and a distributor's one when dealing with divisible commodities. 
We use a game theoretic approach and illustrate several mechanisms to guide the community towards a responsible electricity usage. The model takes into account the preference consumption of every single user and at the same time avoids blackout situations in which the overall consumption is greater than the actual available energy.\\
Our main objective is to select the users' consumption amount in order to optimize and not to waste the produced energy. This aim can be reached through the modification of users' energy consumption according to the produced energy.
The problem that the model faces, is know in literature as an energy management problem, also know as demand side management (DSM) problem or also divisible commodities allocation\cite{dsmp}. It could happen that the available energy for a specific hour of the day is not sufficient to cover users' needs. Instead of providing a larger amount of energy, a distributor is interested into shifting the consumption in the hours in which the overall consumption is much less than the production.\\
The remainder of the paper is organized as follows. Initially, Section \ref{sec:bg} presents the mechanism design background. Then, Section \ref{sec:pro} explain he problem statement as well as the definitions of the proposed mechanisms. In Section \ref{sec:rw}, we expose and analyse related works, while we conclude with remarks and future works in Section \ref{sec:con}.


\section{Background}\label{sec:bg}
\subsection{General concepts}
Mechanism Design \cite{nisan,nara,mas,mt} is based on the concept of social choice that is simply an aggregation of the preferences of the different participants toward a single collective decision. Mechanism Design attempts to implement desired social choices in a strategic setting, assuming that players act rationally in a game theoretic sense. 
\begin{defn}[Player's Valuation Function]
Let us consider a set of players $N = \{1,\dots,n\}$ and a set of \textit{alternatives} or \textit{outcomes} $A$.
Every player $i$ has a preference over alternatives that is described by a \textit{valuation function}: 
\begin{center}
$v_i : A \to \mathbb{R}$
\end{center}
where $v_i(a)$ denotes the valuation that player $i$ assigns to outcome $a$. Furthermore, $v_i \in V_i$ where $V_i \subseteq \mathbb{R}^{|A|}$ is a set of possible valuation functions for player $i$.
\end{defn}
\begin{defn}[Social Choice Function]
The \textit{social choice function} selects an alternative (or outcome) from the set of alternatives $A$ according to the vector of users' valuation functions:
\begin{center}
$f : V_1 \times \dots \times V_n \to A$
\end{center}
So, an outcome $a$, from the set $A$ of alternatives, depends on each possible \textit{profile} $v = (v_1,v_2,\dots,v_n)$ :
\begin{center}
$a = f(v)$
\end{center}
This outcome is called \textit{social choice} for that profile.
\end{defn}
When considering a mechanism with money, that is a mechanism where there are money transfers between the mechanism and players, a payment function computes money transfers for every players. 
\begin{defn}[Direct Revelation Mechanism]
A direct\footnote{There exists also the indirect revelation mechanism with money, that differs for the fact that players have private information (player's preferences) and select strategies according to this information set. In this work, we do not consider indirect revelation mechanism because we assume that the users' valuations functions are known.} revelation mechanism $\mathcal{M}$ is composed of:
\begin{center}
$\mathcal{M} = \langle f, p_1,\dots, p_n \rangle$
\end{center}
where $f$ is the social choice function with $A$ as the possible outcomes and $p_1,\dots, p_n$ are the payment vectors where $p_i : V_1 \times \dots \times \dots V_n \to \mathbb{R}$ is the amount that user $i$ pays to the mechanism.
\end{defn}

\begin{defn}[Utility Function]
Considering a mechanism $\mathcal{M}= \langle f, p_1,\dots, p_n \rangle$, the valuation set $v=(v_1,\dots,v_n)$ and the alternative chosen $a = f(v_1,\dots,v_n)$, the utility for every user $i$ is:
\begin{equation}
u_i(a) = v_i(a) - p_i(v_i(a),v_{-i}(a))
\end{equation}
where $v_{-i}$ is the (n-1)-dimensional vector in which the $i$'th  coordinate is removed.
\end{defn} 
\begin{defn}[Truthfulness]\label{truth}
A mechanism $\mathcal{M}$ is called truthful (also called strategy-proof or incentive compatible) if for every agent $i$, every $v_1 \in V_1, \dots, v_n \in V_n$ and every $v_i' \in V_i$ where $a = f(v_i,v_{-i})$ and $a' = f(v_i',v_{-i})$ then: 
\begin{equation} \label{truth_eq}
u_i(a) \geq u_i(a')
\end{equation}
\end{defn} 
Here, the social choice function selects the outcome $a$ if the player reports his real valuation, at the contrary the social choice function selects the outcome $a'$ if the player, trying to get a greater utility, reports a false valuation. 
In other words, this means that an agent prefers to declare the ``truth" to the mechanism rather than any possible ``lie", because it gives him higher or at least equal utility.
\begin{defn}[Individual Rationality]\label{ir}
A mechanism is individually rational if players always get non-negative utility. Formally if for every $\langle v_1,\dots, v_n \rangle$ we have that: 
\begin{equation}
v_i(f(v_1,\dots, v_n)) - p_i(v_1,\dots, v_n) \geq 0
\end{equation}

\end{defn}

\begin{defn}[No Positive Transfer]\label{npt}
A mechanism has no positive transfers if no player receives money. Formally if for every $v_1,\dots, v_n$ and for every $i$ we have that: 
\begin{equation}
p_i(v_1,\dots, v_n) \geq 0
\end{equation}

\end{defn}
%

\subsection{VCG Mechanism}
\label{sec:vcg}
The most famous direct revelation mechanism is the Vickrey-Clarke-Groves (VCG) Mechanism \cite{nisan}. 
\begin{defn}[VCG Mechanism]
A VCG mechanism determines $f(v)$:
\begin{equation} 
f(v) \in argmax_{b\in A} \sum\limits_{j=1}^{n} v_j(b)
\end{equation}
and $p_i(v)$ such that:
\begin{equation}\label{vcg_p}
p_i(v) = h_i(v_{-i}) - \sum\limits_{j \neq i}^{n} v_j(f(v))
\end{equation}
for some function $h_1,\dots,h_n$ where $h_i : V_{-i} \to \mathbb{R}$. 
\end{defn}
Now, there are several versions of the model according to the choice of $h_i(v_{-i})$. Important to note that the function $h_i$ can be any arbitrary function but it must not depend on the $v_i$. One of the most important versions is the VCG mechanism with the Clarke pivot rule, introduced by Clarke \cite{clarke}.
\begin{defn}[VCG Mechanism with Clarke Pivot Rule]
A VCG mechanism with Clarke pivot payments determines $h_i(v_i)$:
\begin{equation} 
h_i(v_{-i}) = max_a \sum\limits_{j\neq i}^{n} v_j(a)
\end{equation}
so $p_i(v)$ becomes:
\begin{equation}\label{clarke}
p_i(v) = max_b \sum\limits_{j\neq i}^{n} v_j(b) - \sum\limits_{j \neq i}^{n} v_j(f(v))
\end{equation}
where $b$ is the selected alternative if the $i$-th player is not present in the system\footnote{Here, ``$a$" is the optimal assignment including player ``$i$", while ``$b$" is the optimal assignment when we exclude player ``$i$"}.
\end{defn}
By choosing this kind of $h_i(v_{-i})$ Clarke wants to let the buyer paying only the influence that he has in the system. In fact, an user influences the system when the outcome changes depending on the absence or presence of player $i$. If the outcome changes significantly when player $i$ is removed, it means that player $i$ strongly affects the system, so he has to pay for his influence (from this concept derives the name ``Clarke pivot rule").\\
\begin{prop}\label{props}
The VCG mechanism with Clarke pivot rule guaranties the properties of truthfulness, maximize social welfare, individual rationality and no positive transfer \cite{nisan,vcg14}.
\end{prop}

\section{Problem Formulation and System Model}\label{sec:pro}
\subsection{Problem Description}
\begin{figure}
\centering
\begin{subfigure}{.48\textwidth}
  \centering
  \includegraphics[width=1.05\linewidth]{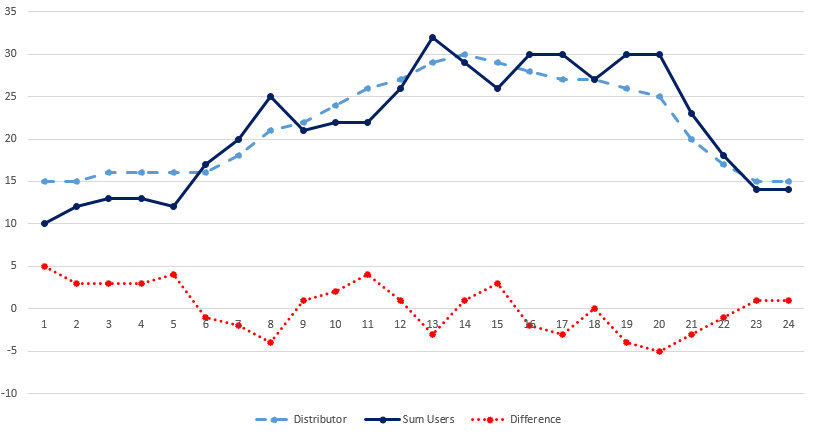}
  \caption{Trend of energy availability considering the energy distributor a player.}
  \label{fig:subgraph2}
\end{subfigure}
\begin{subfigure}{.48\textwidth}
  \centering
  \includegraphics[width=1.05\linewidth]{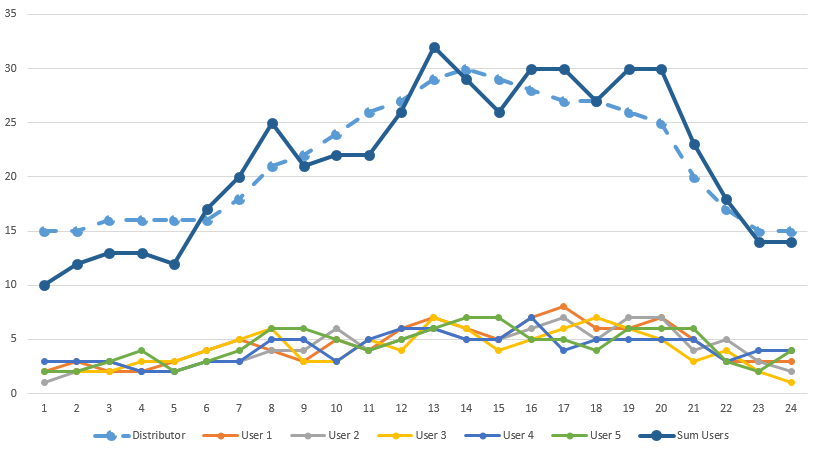}
  \caption{Trend of energy availability considering only consumers as players}
  \label{fig:subgraph1}
\end{subfigure}
\caption{Comparison between energy functions: users' desired trends, the aggregated desired trend of all of agent and the produced energy function.}
\label{fig:graphs}
\end{figure}

In our work, we model an energy allocation problem through a VCG mechanism approach. The main aim is to avoid blackouts when the users' requested energy exceeds the available energy and the energy network must be switched off due to overload. 
Fig. \ref{fig:graphs} describes a possible case for one-day time period with trends for the energy functions. The dashed line represents the distributor's available energy, the continuous line the energy requested from all consumers. The lines on the bottom represent the single consumption of every user (five users in this case). So in Fig. \ref{fig:subgraph1}, we can consider only the consumers "players" and the energy available function as a parameter for the mechanism. The produced energy is a constraint to take into account while maximizing the social welfare.
Fig. \ref{fig:subgraph2} describes an example of a situation in which energy availability function (dashed line) and aggregate users' consumption (continuous line) are compared. Where the set of players contains the distributor and the difference between this two functions (production - consumption, the dotted line in the graph) must be greater or equal than zero. In fact, another way to tackle this problem is that we can consider that the difference between the production and the aggregate consumption, must be greater or equal than zero for this purpose we can consider the distributor as a player. 
The main idea is to deploy a mechanism in order to drive users in shifting energy consumptions, according to the produced energy and the consumption preference of the community.\\

\subsection{Model Description}
Our objective is to determine a mechanism that will select the optimal energy according to the desired consumption of every player. 

\subsubsection{Case 1}\label{sssec:c1}
In our first mechanism we consider a scenario with an energy distributor and $n$ energy users ($n+1$ players), assuming that the $(n+1)^{th}$ is the distributor, so the situation in Fig. \ref{fig:subgraph1}. We have two possible outcomes, $A = \{0,1\}$, the value $1$ represents that the aggregated consumption is not greater than the energy available $P$, the opposite for the value $0$. Every player, except the distributor, has a valuation over the set of alternatives $A$, $v_i : A \to \mathbb{R}$ we define $v_i$ as:
\begin{center}
$v_i(a) = \Bigg\{$ \begin{tabular}{c@{\hskip 30pt}c}
$-x_i$ & if $a = 1$\\
$0$ & if $a = 0$
\end{tabular}
\end{center}
where $x_i$ is the desired consumption of user $i$, assuming that the valuation is directly correlated to the amount of consumed energy as described in \cite{user}. The negative value of the desired consumption is an artefact due to the design of the social choice function and to the fact that we want to model also the energy producer as a player.
In the same way the energy distributor ($n+1^{th}$ player) has a valuation $v_{n+1}$:
\begin{center}
$v_i(a) = \Bigg\{$ \begin{tabular}{c@{\hskip 30pt}c}
$P$ & if $a = 1$\\
$0$ & if $a = 0$
\end{tabular}
\end{center}
where $P$ is the produced energy.
According to the VCG mechanism with pivot rule, the social choice function $f : V \to \{0,1\}$ is defined as:
\begin{equation}
f(v) = \operatorname{arg\,max}_{a \in A} \sum\limits_{j=1}^{n+1} v_j(a)
\end{equation}
The $f(v)$ will select $a = 0$ when overall consumption is greater than the produced energy. 
Otherwise, the $f(v)$ will select $a = 1$ if the overall consumption is less or equal than the produced energy $\sum_{j=1}^{n} v_j(1) \geq -v_{n+1}(1)$. It is possible to note that the mechanism presented guarantees no energy blackout situations, where the produced energy is not sufficient to cover the consumption causing the collapse of the electricity network.\\
As payment scheme we use the Clarke pivot rule (Eq. \ref{clarke}):
\begin{equation}\label{eq_1}
p_i(v) = max_{b \in A} \sum\limits_{j\neq i}^{n+1} v_j(b) - \sum\limits_{j \neq i}^{n+1} v_j(f(v)) 
\end{equation}
Considering a generic user $i$, we note that if $a = 0$ the energy is not provided at all. Furthermore, if $b = 0$, it means that the consumption without user $i$ is greater than the production, user $i$ has not to pay because he is not a pivot player.
However, considering Eq. \ref{eq_1} if $a = 0$ (the energy is not provided) and $b = 1$, it means that if user $i$ is not present in the community the overall consumption is not greater that the produced energy and the energy would be provided. For this reason, user $i$ is pivotal, so he will pay an amount non-negative, corresponding to the energy that would consume other users, even if he has not consumed energy, because it is his fault if the energy is not provided. This, in a real situation, will push user $i$ to not require his allocated energy in order to avoid an heavy charge (and the connected blackout situation).
At the contrary if $a = 1$ the only possible case is to have $b = 1$. In this case users are not pivotal and the mechanism will not charge any additional fee.
This mechanism has the positive aspect that it avoids blackout situations. However, it has a relevant weak point. If the aggregated consumption is greater than the production, the mechanism will not provide energy to anyone even if there could be better solutions by delivering energy only to non-pivot users. 

\subsubsection{Case 2} \label{sssec:c2}
In the previous case, we have two alternatives: everyone or no one consumes energy. This is not an efficient energy management mechanism, because if the consumption is slightly higher than the production, no one is authorized to consume and the whole produced energy is wasted. So, we propose a slightly different mechanism. In this mechanism we have a set of possible outcomes $A \subseteq \{0,1\}^n$ where the value $a[i] = 0$ represents that the specific user $i$ does not consume, the opposite for the value $a[i] = 1$ if the user consumes. 
As described in Fig. \ref{fig:subgraph2}, we consider the consumption of every consumer without the energy distributor but with a threshold value represented by the amount of produced energy. Indeed, we consider the available energy function as a parameter that influences the possible outcome. In fact, we do not consider all the possible elements of the vector $A \subseteq \{0,1\}^n$ but $A = \{a :  \sum_j^n x_j \cdot a[j] \leq P\}$, where $P$ is the energy available and $x_i$ is the desired consumption of player $i$.
In this case, every user has the valuation function $v_i(a) = x_i \cdot a[i]$ (same to Case 1 in \ref{sssec:c1}).\\
Suppose that the total users' consumption is less or equal than produced energy, we have that the social choice function will select the outcome maximizing the sum of valuations. So, the selected alternative is $a = (1,\dots,1)$ with no pivot users. 
In the opposite case, if the consumption is greater than the production, the social choice function selects the alternative $a$ that maximizes the allowed consumption under the production constraint. In other words, it selects to provide energy to a subset of users, according to their optimal consumption, avoiding the energy wasting. Again, considering user $i$, if $a[j] = b[j]$ in Eq.\ref{eq_1} for every $j \neq i$, the outcome is not influenced by user $i$ and his payment remains zero.
But, if $a[j] \neq b[j]$ for some $j \neq i$, it means that without user $i$ the $f(v)$ will choose the consumption of other users in $K$ instead of $i$, with $v_i(a)\geq \sum_{j\in K} v_j(a)$. 
Thanks to this modification, the mechanism presented in this case offers a more efficient energy usage respect to the case in Section \ref{sssec:c1} because here the social choice function can select the maximum number of user that are allowed to consume. In the first case, we had two possibility, provide energy to everyone or not provide energy at all. For this reason, the mechanism could waste all the energy produced. In this second case, the energy is provided only to a subset of users with aim to avoiding the waste of produced energy.
It is still possible to minimize the amount of energy wasted, thanks to a larger set of alternatives $A$ described in the next case.\\

\subsubsection{Case 3} \label{sssec:c3}
In this third case, we propose a mechanism that assigns to users all the available energy till reaching the total amount of produced energy.\\ 
Considering a scenario with $n$ players, we have always a set of possible outcomes $A \subseteq [0,x_i]^n$ where $x_i$ is the desired consumption of user $i$. As in the case in Section \ref{sssec:c2}, we do not consider all the possible elements of $A$ but $A = \{a :  \sum_j^n a[j] \leq P\}$ and $P$ is the energy available. In this way, the distributor can provide an arbitrary amount of energy to user $i$ where the maximum $x_i$ is the $i$'s optimal consumption. The valuation function of user $i$ is: $v_i(a) = a[i]$.\\
Still assuming that ``$a$" is the optimal assignment (outcome selected by the social choice function) including player ``$i$", while ``$b$" is the optimal assignment when we exclude player ``$i$", if $a \neq b$ it means that $a[i] \in (0,x_i]$ and without user $i$ the $f(v)$ will choose the consumption (or assigns a higher consumption) of another user $k$ instead of $i$, resulting that $v_i(a) \geq v_k(a)$. For this reason the payment of user $i$ is non-negative:
\begin{equation}
p_i(v) = \sum\limits_{j\neq i}^{n} v_j(b) - \sum\limits_{j \neq i}^{n} v_j(a) \geq 0  
\end{equation}
The utility of user $i$ will be:
\begin{equation}
u_i(v) = v_i(a) - p_i(v) \geq 0
\end{equation}
In this case, we can distinguish two cases. Let $C_{tot} = \sum_{i\neq j} x_j$ (the sum of all requests of all users except user $i$).
\begin{enumerate}
\item if $C_{tot} \geq P$, then $p_i = a[i]$ meaning that a user $i$ will pay his consumed energy. In fact we have that $p_i = \sum v_j(b) - \sum v_j(a)$ where $\sum v_j(b) = P$, while $\sum v_j(a) = P-a[i]$;
\item if $C_{tot} < P$, then $p_i = C_{tot} - (P-a[i])$, because again $p_i = \sum v_j(b) - \sum v_j(a)$, but here $\sum v_j(b)=C_{tot}$. This means that $p_i$ can be 0 also if $a[i]=P-C_{tot}$, which can be $> 0$.
\end{enumerate}
It is important to note that every user $i$ will receive an amount of energy not greater than $x_i$ because of the constraint: $a[i] \in [0,x_i]$.\\
This mechanism has several properties: 
\begin{itemize}
\item the whole produced energy is used. Indeed, the social choice function selects an outcome $a$ which has the property: $\sum_i^n a[i] = P$. This is an optimal solution because there is no waste of energy;
\item the mechanism has an infinite set of possible solutions (choices of $a$), all those that satisfy the production threshold; 
\item as in the case in Section \ref{sssec:c2}, every player will get in every case a non-negative utility.
\end{itemize}
The models analysed so far consider only one slice of time. A natural improvement in the next case is the insertion of the concept of time.\\

\subsubsection{Case 4} \label{sssec:c4}
We start from the mechanism described in Section \ref{sssec:c2}. But, the main difference is that in this fourth case we take into account also the time $t$. Formally, we introduce a time variable: $t \in [0,T]$, where $T$ is the maximum length.\\ 
Considering a scenario with $n$ players, we have that a possible outcome is $a = (a_1(t),\dots,a_n(t))$ where $a_i(t) \in \mathbb{R}^+$ for every $t \in [0,T]$ and where $x_i : [0,T] \to \mathbb{R}$ is the desired consumption function of user $i$ over time $t$. The function $a_i(t)$ represents the assigned power to user $i$ for every tick $t$, in the same way the function $P(t)$ is the function of available power for every tick $t$, where the energy is the power per unit time.\\
The valuation function is: 
\begin{center}
$v_i(a) = \Bigg\{$ \begin{tabular}{c@{\hskip 30pt}c}
$a_i(t)$ & if $a_i(t) \leq x_i(t)$\\
$x_i(t)$ & if $a_i(t) > x_i(t)$
\end{tabular}
\end{center}
representing the total amount of energy received by the i-th customer once the excess energy has been discarded.
According to the VCG mechanism with pivot rule, the social choice function $f : V \to \{0,1\}$ is defined as:
\begin{equation}
f(v) = \operatorname{arg\,max}_{a \in A} \sum\limits_{j=1}^{n} v_j(a)
\end{equation}
As in the case in Section \ref{sssec:c3}, we do not consider all the possible elements of $\mathbb{R}^n$, in fact $A$ is defined as:
\begin{equation}
A = \bigg\{a :  \sum\limits_{j =1 }^{n} a_i(t) \leq P(t), \forall t \in [0,T] \;\;\; \bigwedge \;\;\; \int\limits_{0}^{T} x_i(t) dt \leq \int\limits_{0}^{T} a_i(t) dt \; \forall i \in \{1,\dots,n\} \bigg\}
\end{equation}
The first condition on $a$ ensures that for every $t$ the overall consumption is not greater than the produced power. Then, the second condition establishes that the amount of energy provided to user $i$ is greater or equal to his optimal amount of requested energy. With this second constraint we want to shift the users' consumption instead of not providing power at all.\\
If the overall consumption $x(t) = \sum_{j=1}^{n} x_j(t)$ is not greater that the production $P(t)$ for every $t$, it means that the $f(v)$ will choose $a(t) = x(t)$, and there are no pivot users so $p_i(v) = 0$ for every user $i$. Consequently, the utility of user $i$ will be: $u_i(v) = a_i(t)$.\\
If for some $t$, $x(t) > P(t)$ the social choice function will select an outcome $a$ that for that $t$ will be $\sum_{j=1}^{n} a_j(t) = P(t)$, providing less power to one or more users but providing more power to this set of users where $x(t) < P(t)$ in order to satisfy the constraint: $\int_{0}^{T} x_i(t)dt \geq \int_{0}^{T} a_i(t)dt$.\\ 
This mechanism has all the properties of the mechanism in Section \ref{sssec:c4}, in addition it chooses the energy to be provided according to every user's preferences, allowing the shifting of the consumption in the interval $[0,T]$.

\subsubsection{Case 5} \label{sssec:c5}
In the previous case, it can happen that there is no solution because is not possible to shift the consumption for every user because it is not satisfied the second constraint regarding the single consumption for each user.
So, in this new case we try to overcome this problem. Our aim is to allocate a positive amount of power for users that have a positive consumption. To reach this aim, we modify the set of possible outcomes $A$. As in the mechanism \ref{sssec:c4}, we have that a possible outcome is $a = (a_1(t),\dots,a_n(t))$ where $a_i(t) \in [0,x_i(t)]$ for every $t \in [0,T]$ and where $x_i : [0,T] \to \mathbb{R}$ is the desired consumption function of user $i$ over time $t$. The allocated power function for every user $i$ is defined as:
\begin{center}
$a_i(t) = \Bigg\{$ \begin{tabular}{c@{\hskip 30pt}c}
$x_i(t)$ & if $\sum\limits_{j=1}^{n} x_j(t)\leq P(t)$\\
$\frac{x_i(t)}{\sum\limits_{i=1}^{n} x_i(t)}\cdot P(t)$ & if $\sum\limits_{j=1}^{n} x_j(t) > P(t)$
\end{tabular}
\end{center}
As payment scheme we use the Clarke pivot rule (Equation \ref{clarke}):
\begin{equation}
p_i(v) = max_{b \in A} \sum\limits_{j\neq i}^{n+1} v_j(b) - \sum\limits_{j \neq i}^{n+1} v_j(f(v)) 
\end{equation}
According to this allocation scheme, if the requested power is less or equal than the production the mechanism will select the users' desired consumption. In the opposite case, the mechanism will allocate a power amount proportional to the $i$'s desired consumption.
This approach has the advantage that every user will get a positive amount of power and no energy is wasted.

\subsubsection{Case 6} \label{sssec:c6}
We start from the mechanism described in Section \ref{sssec:c4}.
So, consider a scenario with $n$ players, we have that a possible outcome is $a = (a_1(t),\dots,a_n(t))$ where $a_i(t) \in \mathbb{R}^+$ for every $t \in [0,T]$ and where $x_i : [0,T] \to \mathbb{R}$ is the desired consumption function of user $i$ over time $t$. The function $a_i(t)$ represents the assigned power to user $i$ for every tick $t$, in the same way the function $P(t)$ is the function of available power for every tick $t$.\\
The valuation function is: 
\begin{center}
$v_i(a) = \Bigg\{$ \begin{tabular}{c@{\hskip 30pt}c}
$a_i(t)$ & if $a_i(t) \leq x_i(t)$\\
$x_i(t)$ & if $a_i(t) > x_i(t)$
\end{tabular}
\end{center}
According to the VCG mechanism with pivot rule, the social choice function $f : V \to \{0,1\}$ is defined as:
\begin{equation}
f(v) = \operatorname{arg\,max}_{a \in A} \sum\limits_{j=1}^{n} v_j(a)
\end{equation}
As in the case in Section \ref{sssec:c3}, we do not consider all the possible elements of $A$, in fact $A$ is defined as:
\begin{equation}
A = \bigg\{a :  \sum\limits_{j =1 }^{n} a_i(t) \leq P(t), \forall t \in [0,T] \;\;\; \bigwedge \;\;\; \int\limits_{0}^{T} x_i(t) dt \geq \int\limits_{0}^{T} a_i(t) dt \bigg\}
\end{equation}
The payment made by every user $i$ is proportional to his consumption plus a positive amount if the provided power is near the produced power function. Formally, it is defined as: 
\begin{equation}
p_i(v) = a_i(t) + \alpha(y)
\end{equation}
where $\alpha : R^+ \to R^+$ and $y = \big[ \sum_j^n a_j(t) - P(t)\cdot c \big]^+$ with $c \in (0,1]$. The value $c$ is the percentage of produced power which ensures that there will not be a blackout situation. If for a specific $t$ we exceed this safe amount every user has to pay an extra amount of money, because the risk of having a blackout increases. 

\subsection{Further Remarks}
We could sum up our problem by drawing connections with the classical knapsack optimization problem. In Case \ref{sssec:c1}, we decide to put all into the knapsack only if the volume of the knapsack is sufficient, otherwise we do not put in anything. In the energy case, the distributor will provide energy only if the consumption is not greater than the available energy. In Case \ref{sssec:c2}, we assume that we have a set of object to put into the knapsack minimizing the empty space. So, the distributor will provide energy to a subset of users. In Case \ref{sssec:c3}, we assume that we have a liquid instead of a set of objects, so we are able to fill the entire knapsack. Here as well, the mechanism selects the consumption for a subset of users, however there is no wasted energy. In Case \ref{sssec:c4}, we introduce the time variable. So, every energy function become a power function over time. In Case \ref{sssec:c5}, we assign the energy in a proportional way to users maintining the VCG payment scheme. In Case \ref{sssec:c6}, we apply a mechanism without the VCG payment rule, but a user will pay according to the proximity to the produced energy threshold.
A final remark is that this model is essentially a game so players must be motivated to play getting a positive utility. But, in our case study, energy allocation, a user usually has to consume and, consequently, play the game for this reason he can accept also an utility equal to zero.




\section{Related Work}\label{sec:rw}
In the last years, the mechanism design is used to solve several kinds of problems, thanks to its ability of solving optimization issue in a distributed manner. The application fields are various, starting from economic markets to energy efficiency. In \cite{wsn}, a mechanism is applied to the problem of allocating resources into wireless networks. The problem is modelled as a strategic game with a resource pricing scheme where the prices are decided by a mechanism designer. Taking into account prices and local preferences, wireless devices decide which action to perform. The final aim is to maximize social welfare and energy efficiency.\\
Work \cite{cloud} involves real users (and not machines) into the mechanism. The aim is to select the best price for cloud services. Here, different scenarios are evaluated, where users access the service for the same time-period and where users come and go. This work applies the Shapley Value mechanism.\\
An approach close to our work from the state of art is represented by the paper \cite{dsm2}, and its related previous work \cite{opti}.
In \cite{dsm2}, the authors associate the users' needs with energy producers. 
The aim of the model is to encourage efficient energy consumption among users with a VCG mechanism that collects private informations from users and computes the electricity bill. In this work, users receive all the required energy because the energy provider is able to buy energy from external sources. The main difference is that we assume that there exists a threshold of produced energy that users cannot overcome, i.e. there are some situations in which is not possible to buy or produce more electricity. So, the efficiency of our model is computed through the maximization of users’ valuations. Instead, in \cite{dsm2} they want to maximize the users’ valuations while minimizing the total energy cost paid by the provider to the producer. In summary, we assign the available energy to users. So, we face a resource allocation problem instead of supply the overall requested energy at the best price.
Another similar work is reported in \cite{drp}, that tries to achieve efficiency in energy consumption with economic incentives, thanks to an indirect revelation mechanism.
A survey of residential load controlling techniques to implement demand side management in future smart grid is presented in \cite{survey}. In that work, the authors discuss different energy consumption scheduling schemes to minimize energy consumption cost and reduce peak load. They analyse several methods based on different approaches, such as scheduling, game theory or scheduling protocol trough smart meter devices.
The paper \cite{dsm} presents a new energy consumption scheduling scheme to enable Demand Side Management (DSM) in which an autonomous energy scheduling scheme is deployed to achieve minimum consumption cost and reduction in peak load by the use of an algorithm taking into account peak-hours with smart meter devices.
In several works, the VCG mechanism is modified to solve allocation problems. A concrete example is described in \cite{kelly}, in which a classical VCG mechanism is combined with one-dimensional surrogate valuation functions. This approach provides efficient allocation for strategic buyers at Nash equilibrium points.
The work in \cite{envy} presents a pseudo-VCG mechanism, based on the mechanism in \cite{kelly}. In \cite{envy}, the authors present a mechanism for allocating divisible commodity, satisfying several properties as the No Envy property. In \cite{resallo}, several auction mechanisms are evaluated for network resource allocation without knowledge of the utility functions of involved actors. It deals with maximum allocation efficiency for one or multiple indivisible items and for divisible resources applying several auction mechanism. 
Considering users' energy behaviour, every lifestyle is strictly related to energy absorption. In the literature, several works have been published in the topic of users' energy behaviour, trying to classify different user's profile. In particular, in \cite{user} there is a significant study of patterns of domestic electricity consumption. This work is based on a real analysis of different houses taking into account several parameters like: location, floor area, household, etc. This paper states that there is a direct correlation between the amount of energy consumed and the user's well-being. Moreover, we assume to have the users' consumption function.\\

\section{Conclusion and Future Work}\label{sec:con}
In this paper, we applied a VCG mechanism for the DSM problem. We propose several configurations of the mechanism, starting from the simplest to a more complicated configuration which has the property of assigning the available energy to users according to their desired energy minimizing the energy wasting while maximizing the aggregate utility of all users. 
One weak point of the mechanisms presented is that users won't pay according to the amount of energy assigne to them but they pay a different amount with respect to their energy consumption. This fact collides with the idea behind mechanism design: stimulate the players by offering an advantage in playing the game. For this reason, the player usually gets a positive utility. However, in our specific problem users need to consume energy and consequently play the game. So, a development is to find a different payment scheme that takes into account the actual consumption and energy consumption peaks. The final aim is to stimulate users to behave in a good energy way offering a discount on the electricity bill that will lead to get a positive net utility. In a simulation context, it will be extremely useful to use real consumption functions and a real energy availability function in order to build a multiagent simulation system with the final aim to capture and compare the results between standard and our proposed mechanism.



\bibliographystyle{splncs03}
\bibliography{article_20160825}

\def\germ{\frak} \def\scr{\cal} \ifx\documentclass\undefinedcs
  \def\bf{\fam\bffam\tenbf}\def\rm{\fam0\tenrm}\fi 
  \def\defaultdefine#1#2{\expandafter\ifx\csname#1\endcsname\relax
  \expandafter\def\csname#1\endcsname{#2}\fi} \defaultdefine{Bbb}{\bf}
  \defaultdefine{frak}{\bf} \defaultdefine{=}{\B} 
  \defaultdefine{mathfrak}{\frak} \defaultdefine{mathbb}{\bf}
  \defaultdefine{mathcal}{\cal}
  \defaultdefine{beth}{BETH}\defaultdefine{cal}{\bf} \def\bbfI{{\Bbb I}}
  \def\mbox{\hbox} \def\text{\hbox} \def\om{\omega} \def\Cal#1{{\bf #1}}
  \def\pcf{pcf} \defaultdefine{cf}{cf} \defaultdefine{reals}{{\Bbb R}}
  \defaultdefine{real}{{\Bbb R}} \def\restriction{{|}} \def\club{CLUB}
  \def\w{\omega} \def\exist{\exists} \def\se{{\germ se}} \def\bb{{\bf b}}
  \def\equivalence{\equiv} \let\lt< \let\gt>
\begin{thebibliography}{10}
\providecommand{\url}[1]{\texttt{#1}}
\providecommand{\urlprefix}{URL }

\bibitem{drp}
Barreto, C., Mojica-Nava, E., Quijano, N.: Design of mechanisms for demand
  response programs. In: 52nd IEEE Conference on Decision and Control. pp.
  1828--1833 (Dec 2013)

\bibitem{vcg14}
Blum, A.: Algorithms, games, and networks (2013),
  \url{http://www.cs.cmu.edu/~arielpro/15896/docs/notes14.pdf}, lectures notes
  14

\bibitem{wsn}
Chorppath, A.K., Alpcan, T.: Mechanism design for energy efficiency in wireless
  networks. In: Modeling and Optimization in Mobile, Ad Hoc and Wireless
  Networks (WiOpt), 2011 International Symposium on. pp. 389--394. IEEE (2011)

\bibitem{clarke}
Clarke, E.H.: Multipart pricing of public goods. Public Choice  11(1),  17--33
  (1971), \url{http://dx.doi.org/10.1007/BF01726210}

\bibitem{dsmp}
Gellings, C.: The concept of demand-side management for electric utilities.
  Proceedings of the IEEE  73(10),  1468--1470 (Oct 1985), iSSN:0018-9219,
  DOI:10.1109/PROC.1985.13318

\bibitem{user}
Ghaemi, S., Brauner, G.: User behavior and patterns of electricity use for
  energy saving. Internationale Energiewirtschaftstagung an der TU Wien, IEWT
  (2009)

\bibitem{mt}
Jackson, M.O.: Mechanism theory (2001)

\bibitem{resallo}
Koutsopoulos, I., Iosifidis, G.: Auction mechanisms for network resource
  allocation. In: Modeling and Optimization in Mobile, Ad Hoc and Wireless
  Networks (WiOpt), 2010 Proceedings of the 8th International Symposium on. pp.
  554--563. IEEE (May 2010)

\bibitem{dsm}
Mahmood, A., Ullah, M., Razzaq, S., Basit, A., Mustafa, U., Naeem, M., Javaid,
  N.: A new scheme for demand side management in future smart grid networks.
  Procedia Computer Science  32(Complete),  477--484 (2014)

\bibitem{nara}
Narahari, Y.: Game Theory and Mechanism Design, chap.~14. World Scientific
  Publishing Company Pte. Limited (2014)

\bibitem{nisan}
Nisan, N., Roughgarden, T., Tardos, E., Vazirani, V.V.: Algorithmic Game
  Theory, chap.~9. Cambridge University Press (2007)

\bibitem{dsm2}
Samadi, P., Mohsenian-Rad, H., Schober, R., Wong, V.W.S.: Advanced demand side
  management for the future smart grid using mechanism design. IEEE
  Transactions on Smart Grid  3(3),  1170--1180 (Sept 2012)

\bibitem{opti}
Samadi, P., Schober, R., Wong, V.W.S.: Optimal energy consumption scheduling
  using mechanism design for the future smart grid. In: SmartGridComm. pp.
  369--374. IEEE (2011),
  \url{http://dblp.uni-trier.de/db/conf/smartgridcomm/smartgridcomm2011.html#SamadiSW11}

\bibitem{mas}
Shoham, Y., Leyton-Brown, K.: Multiagent Systems: Algorithmic, Game-Theoretic,
  and Logical Foundations, chap.~10. Cambridge University Press, New York, NY,
  USA (2008)

\bibitem{survey}
Ullah, M.N., Mahmood, A., Razzaq, S., Ilahi, M., Khan, R.D., Javaid, N.: A
  survey of different residential energy consumption controlling techniques for
  autonomous {DSM} in future smart grid communications. CoRR  abs/1306.1134
  (2013), \url{http://arxiv.org/abs/1306.1134}

\bibitem{cloud}
Upadhyaya, P., Balazinska, M., Suciu, D.: How to price shared optimizations in
  the cloud. Proc. VLDB Endow.  5(6),  562--573 (Feb 2012),
  \url{http://dx.doi.org/10.14778/2168651.2168657}

\bibitem{kelly}
Yang, S., Hajek, B.: Vcg-kelly mechanisms for allocation of divisible goods:
  Adapting vcg mechanisms to one-dimensional signals. In: 2006 40th Annual
  Conference on Information Sciences and Systems. pp. 1391--1396. IEEE (March
  2006)

\bibitem{envy}
You, J.S.: Envy-free and incentive compatible division of a commodity. R\&R,
  Economic Letters  (2015)

\end{thebibliography}

\end{document}